\documentclass[prb,twocolumn,showpacs,bm,amssymb,amsmath,floatfix]{revtex4}
\usepackage{epsfig}
\begin{document}
\title{Equivalence of the Falicov-Kimball and Brandt-Mielsch forms for the
free energy of the infinite-dimensional Falicov-Kimball model}
\date{\today}
\author{A.~M.~Shvaika}
\email{ashv@icmp.lviv.ua} 
\homepage{http://ph.icmp.lviv.ua/~ashv}
\affiliation{Institute for Condensed Matter Physics of the National
Academy of Sciences of Ukraine, 1 Svientsitskii Str., 79011 Lviv,
Ukraine}
\author{J.~K.~Freericks}
\email{freericks@physics.georgetown.edu}
\homepage{http://www.physics.georgetown.edu/~jkf}
\affiliation{Department of Physics, Georgetown University, Washington, DC
20057}

\begin{abstract}
Falicov and Kimball proposed a real-axis
form for the free energy of the Falicov-Kimball
model that was modified for the coherent potential approximation by Plischke.
Brandt and Mielsch proposed an imaginary-axis form for the free energy of
the dynamical mean field theory solution of the Falicov-Kimball
model.  It has long been known that these
two formulae are numerically equal to each other; an explicit derivation
showing this equivalence is presented here.
\end{abstract}


\maketitle

The Falicov-Kimball model\cite{falicov_kimball_1969} is one of the simplest
many-body Hamiltonians.  It was introduced in
1969 to describe metal-insulator transitions in a number of rare-earth
and transition-metal compounds and was solved in the limit of infinite
dimensions by Brandt and Mielsch\cite{brandt_mielsch_1989,brandt_mielsch_1990,%
brandt_mielsch_1991}.  The earlier work of Falicov's
group\cite{ramirez_falicov_kimball_1970} was modified by
Plischke\cite{plischke_1972} for the coherent-potential approximation to
give an explicit formula for the Helmholz free energy in terms of integrals
over the interacting density of states (DOS).  Later, Brandt and
Mielsch\cite{brandt_mielsch_1991} derived an exact formula for the Helmholz
free energy in terms of summations over Matsubara frequencies in the
infinite-dimensional limit.  Numerical evaluation of these two forms for the
free energy showed that they were indeed equal\cite{chung_freericks_1998,%
chung_freericks_2000} but no explicit derivation of the equivalence has
appeared.

We illustrate this equivalence here for the spinless version of the
Falicov-Kimball model (generalization to higher-spin versions is simple).
The spinless Falicov-Kimball Hamiltonian\cite{falicov_kimball_1969} is
\begin{equation}
\mathcal{H}=-\sum_{ij}t_{ij}c^\dagger_ic_j+E_f\sum_if^\dagger_if_i+U
\sum_ic^\dagger_ic_if^\dagger_if_i
\label{eq: ham}
\end{equation}
where $c^\dagger_i$ ($c_i$) creates (destroys) an itinerant electron at site
$i$, $f^\dagger_i$ ($f_i$) creates (destroys) a localized electron at site
$i$, $t_{ij}$ is the Hermitian hopping matrix (which is chosen to be
nonzero only between nearest neighbors), $E_f$ is the localized electron
site energy, and $U$ is the on-site Coulomb interaction between localized
and itinerant electrons.  Chemical potentials $\mu$ and $\mu_f$ are employed
for the intinerant and localized electrons, respectively.

In the limit where the spatial dimension $d$ becomes large, the many-body
problem can be solved exactly when the hopping is chosen to
scale\cite{metzner_vollhardt_1989} as $t=t^*/2\sqrt{d}$.  In this case,
the so-called local approximation becomes exact.  We sketch the algorithm
used to solve the many-body problem, in order to establish our notation.

The local Green's function $G(z)$ can be written as the Hilbert transform of
the noninteracting DOS $\rho(\epsilon)$
\begin{equation}
G(z)=\int d\epsilon\rho(\epsilon)\frac{1}{z+\mu-\Sigma(z)-\epsilon}
\label{eq: g_hilbert}
\end{equation}
with $z$ in the complex plane and $\Sigma(z)$ the local self energy.
Dyson's equation for the local self energy reads
\begin{equation}
\Sigma(z)=z+\mu-\lambda(z)-G^{-1}(z)
\label{eq: dyson}
\end{equation}
with $\lambda(z)$ the dynamical mean field (which must be determined self
consistently).  Solving the atomic problem in a time-dependent field yields
another equation for the local Green's function
\begin{equation}
G(z)=\frac{w_0}{z+\mu-\lambda(z)}+\frac{w_1}{z+\mu-\lambda(z)-U}
\label{eq: g_atom}
\end{equation}
with $w_0=\mathcal{Z}_0/\mathcal{Z}$, $w_1=\mathcal{Z}_1/\mathcal{Z}$
(the localized electron density),
and $\mathcal{Z}=\mathcal{Z}_0+\mathcal{Z}_1$ (the atomic partition
function).  The symbols $\mathcal{Z}_0$
and $\mathcal{Z}_1$ can be expressed as infinite products
\begin{equation}
\mathcal{Z}_0=(1+e^{\beta\mu})\prod_n\left ( 1-\frac{\lambda_n}{i\omega_n+\mu}
\right )
\label{eq: z0_def}
\end{equation}
and
\begin{equation}
\mathcal{Z}_1=e^{-\beta(E_f-\mu_f)}
(1+e^{\beta(\mu-U)})\prod_n\left ( 1-\frac{\lambda_n}{i\omega_n+\mu-U}
\right )
\label{eq: z1_def}
\end{equation}
where $\beta=1/T$, and we used the notation $\lambda_n=\lambda(i\omega_n)$
with $i\omega_n=i\pi T(2n+1)$ the fermionic Matsubara frequency.

\begin{figure}[htb]
\epsfxsize=3.0in
\epsffile{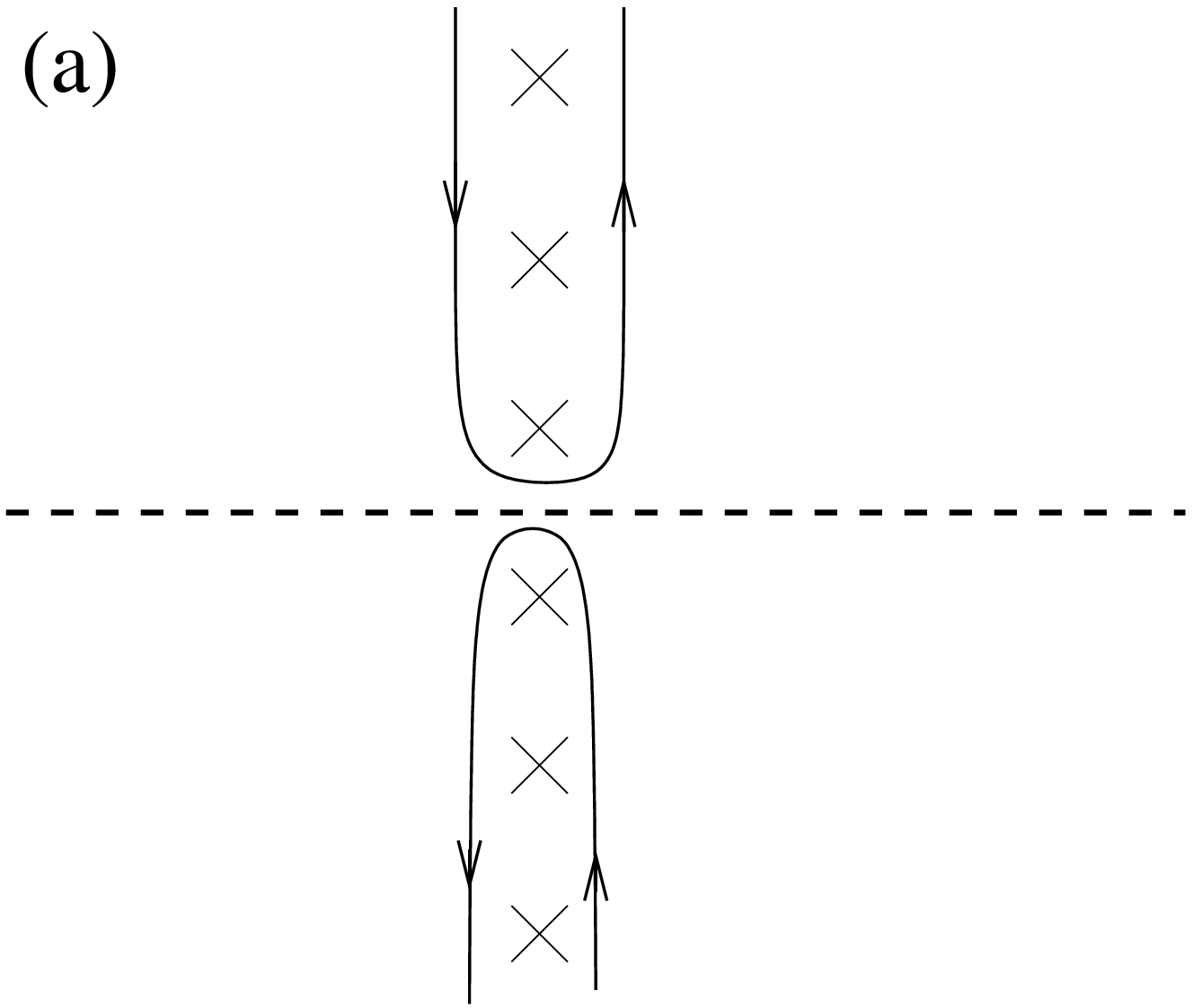}
\epsfxsize=3.0in
\epsffile{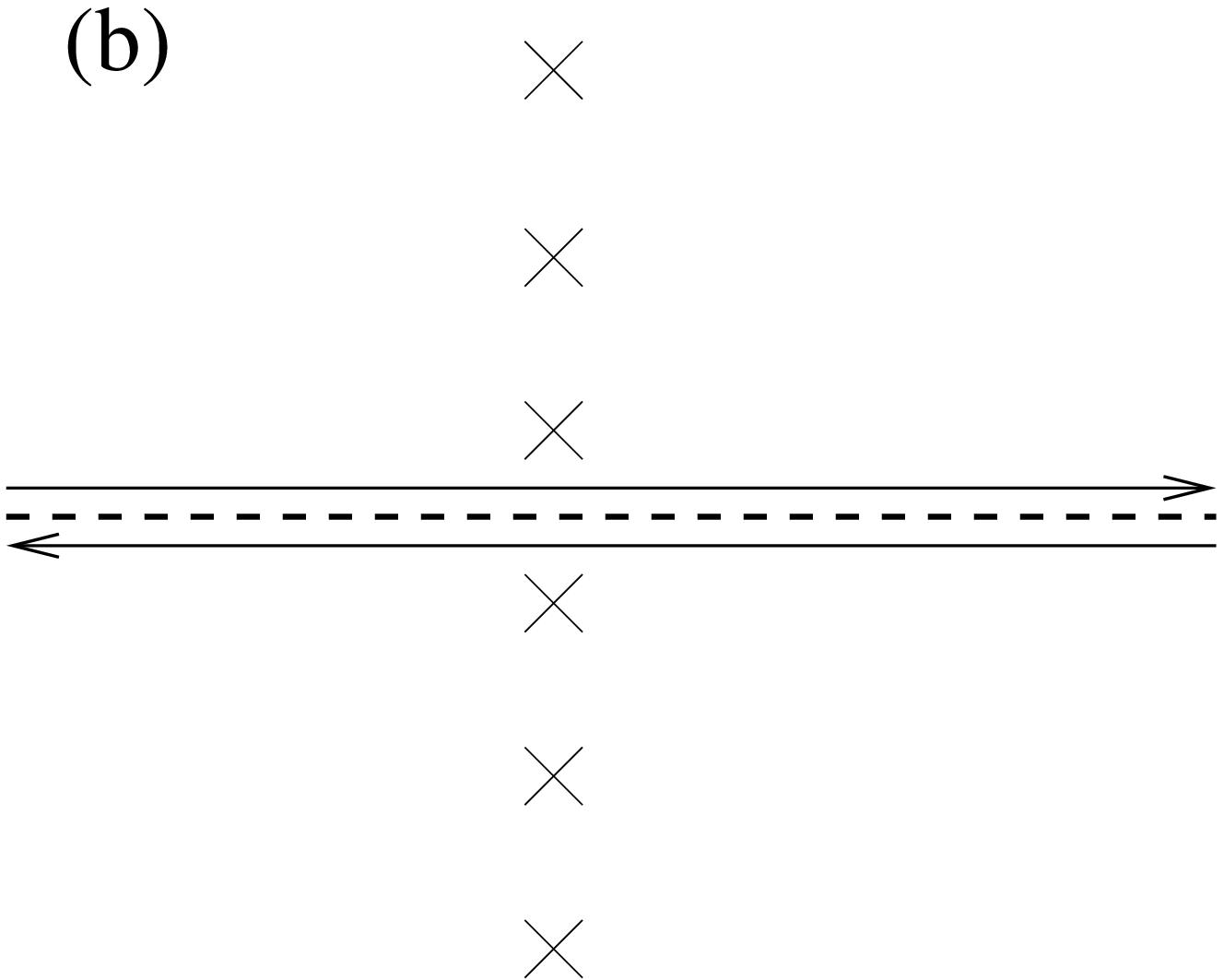}
\caption{Contours used in various integrals.  In panel (a), we show the
contour $C$ that surrounds all of the fermionic Matsubara frequencies
which are indicated by X's.  The dotted line denotes the real axis.
In panel (b), we show the deformed contour $C^\prime$ that allows one
to replace the integral by one over the real axis.
\label{fig: contour}}
\end{figure}

The Brandt-Mielsch form for the Helmholz free energy is
\begin{eqnarray}
\mathcal{F}&=&-T\ln\mathcal{Z}-T\int d\epsilon \rho(\epsilon)\sum_n\ln[(i\omega_n
+\mu-\Sigma_n-\epsilon)G_n]\cr
&+&\mu_fw_1+\mu\rho_c
\label{eq: free_bm}
\end{eqnarray}
with $\rho_c$ the itinerant electron density. Our aim is to replace the
Matsubara frequency summation of the logarithmic function by an integral
over the real axis. To do this we use Eq.~(\ref{eq: dyson}) to write
$G_n=1/(i\omega_n+\mu-\lambda_n-\Sigma_n)$ and rewrite the sum in %
Eq.~(\ref{eq: free_bm}) as
\begin{eqnarray}
&~&T\sum_n \ln [(i\omega_n+\mu-\Sigma_n-\epsilon)G_n]=\cr
  &~&T\sum_n\Biggr[\ln\left(1-\frac{\epsilon}{i\omega_n+\mu-\Sigma_n}\right)\cr
&~&
  -\ln\left(1-\frac{\lambda_n}{i\omega_n+\mu-\Sigma_n}\right)\Biggr].
\label{eq: free_bm1}
\end{eqnarray}
The function $\Xi(z)=1/[z+\mu-\Sigma(z)]$ is the irreducible part (with respect
to the hopping) of the itinerant electron 
Green's function and it possesses the same
analytic properties as do the Green's functions
(a branch cut on the real axis with a change in sign of the imaginary part
above or below the cut\cite{shvaika_2002b}). The dynamical mean field 
$\lambda(z)$ also has the same analytic properties. As a result, the logarithmic
functions in Eq.~(\ref{eq: free_bm1}) are analytic functions above and
below the real axis (the only branch cut lies on the real axis) and they
behave as $1/z$ for $|z|\to\infty$. This implies that
we can express the Matsubara
frequency summation as a contour integral around the contour $C$
illustrated in Fig.~\ref{fig: contour}~(a) yielding
\begin{eqnarray}
&~&T\sum_n \ln [(i\omega_n+\mu-\Sigma_n-\epsilon)G_n]=\cr
 &~&\frac{1}{2\pi i}\int_C dz f(z)
 \Biggl[\ln\left(1-\frac{\epsilon}{z+\mu-\Sigma(z)}\right)\cr
  &~&-\ln\left(1-\frac{\lambda(z)}{z+\mu-\Sigma(z)}\right)\Biggr]
\label{eq: contour1}
\end{eqnarray}
with $f(z)=1/[1+\exp(\beta z)]$ the Fermi-Dirac distribution. The contour
$C$ is deformed to $C^\prime$ which runs parallel to the real axis as
shown in Fig.~\ref{fig: contour}~(b).  Since there is a branch cut on the real
axis, the integral over $C^\prime$ becomes the imaginary part of the
integral from $-\infty$ to $\infty$  
\begin{eqnarray}
 \mathcal{F}&=&-T\ln\mathcal{Z}+\mu_fw_1+\mu\rho_c\cr
 &+&\frac{1}{\pi}\int d\omega\int d\epsilon
 \rho(\epsilon)f(\omega)
 \Biggl[\textrm{Im}\ln\left(1-\frac{\epsilon}{\omega+\mu-\Sigma(\omega)}\right)\cr
 && -\textrm{Im}\ln\left(1-\frac{\lambda(\omega)}{\omega+\mu-\Sigma(\omega)}\right)\Biggr].
 \label{eq: contour1a}
\end{eqnarray}
Because the sign of the imaginary part of the functions that make up the
argument of the logarithms
is fixed above and below the real axis, the value of the imaginary part of the
logarithms is defined to lie in the range between $-\pi$ and $0$ or $0$ and
$\pi$, depending on this sign. To satisfy the analytic properties of the
logarithms in Eq.~(\ref{eq: contour1a}), note that the expression
in the square brackets can be rewritten as
\begin{equation}
  \textrm{Im}\ln\frac{\omega+\mu-\Sigma(\omega)-\epsilon}
                     {\omega+\mu-\Sigma(\omega)-\lambda(\omega)},
\end{equation}
but one must be careful not to shift the imaginary part of the 
logarithm by an integer multiple of $2\pi$, which corresponds to a different
sheet of the logarithm.

Noting that
\begin{equation}
f(\omega)=-T\frac{d}{d\omega} \ln [1+\exp(-\beta\omega)],
\label{eq: f_deriv}
\end{equation}
allows us
to integrate by parts (since the boundary terms vanish) and gives
\begin{eqnarray}
\mathcal{F}&=&-T\ln\mathcal{Z}+\mu_fw_1+\mu\rho_c\cr
&+&\frac{T}{\pi}\int d\omega\int d\epsilon
\rho(\epsilon)\ln [1+e^{-\beta\omega}]\cr
&\times&\textrm{Im}\left [ \frac{1-\Sigma^\prime(\omega)}
{\omega+\mu-\Sigma(\omega)-\epsilon}-\frac{1-\Sigma^\prime(\omega)-
\lambda^\prime(\omega)}{\omega+\mu-\lambda(\omega)-\Sigma(\omega)}\right ]
\label{eq: contour2}
\end{eqnarray}
with the prime indicating a derivative with respect to $\omega$.  The integral
over $\epsilon$ can be performed by using Eq.~(\ref{eq: g_hilbert}) and
the fact that the DOS has unit weight, to yield
\begin{eqnarray}
\mathcal{F}&=&-T\ln\mathcal{Z}+\mu_fw_1+\mu\rho_c\cr
&+&\frac{T}{\pi}\int d\omega
\ln [1+e^{-\beta\omega}]\textrm{Im}[G(\omega)\lambda^\prime (\omega)].
\label{eq: contour3}
\end{eqnarray}
The interacting DOS is defined to be $A(\omega)=-\textrm{Im}[G(\omega)]/\pi$.
Using this fact, we can add and subtract an integral over $A(\omega)$ to
produce
\begin{eqnarray}
\mathcal{F}&=&-T\int d\omega A(\omega)\ln(1+e^{-\beta\omega})
-T\ln\mathcal{Z}+\mu_fw_1+\mu\rho_c\cr
&+&\frac{T}{\pi}\int d\omega
\ln [1+e^{-\beta\omega}]\textrm{Im}[G(\omega)\{-1+\lambda^\prime (\omega)\}].
\label{eq: contour4}
\end{eqnarray}
Next, we substitue in Eq.~(\ref{eq: g_atom}) for $G(\omega)$ and add
\begin{eqnarray}
0&=&\frac{T}{\pi}\int d\omega\ln (1+e^{-\beta\omega})\textrm {Im}\left [
\frac{w_1}{\omega+\mu-U+i0^+}\right ]\cr
&+&Tw_1\ln(1+e^{\beta(\mu-U)})
\label{eq: dsum1}
\end{eqnarray}
and
\begin{eqnarray}
0&=&\frac{T}{\pi}\int d\omega\ln (1+e^{-\beta\omega})\textrm {Im}\left [
\frac{1-w_1}{\omega+\mu+i0^+}\right ]\cr
&+&T(1-w_1)\ln(1+e^{\beta\mu})
\label{eq: dsum2}
\end{eqnarray}
to Eq.~(\ref{eq: contour4}). Collecting terms gives
\begin{eqnarray}
\mathcal{F}&=&-T\int d\omega A(\omega)\ln(1+e^{-\beta\omega})
-T\ln\mathcal{Z}+\mu_fw_1+\mu\rho_c\cr
&+&\frac{T}{\pi}\int d\omega
\ln [1+e^{-\beta\omega}]\cr
&\times&\textrm{Im}\Biggr \{ \frac{w_1}{\omega+\mu-U+i0^+}\left [ 1+
\frac{(\omega+\mu-U)[-1+\lambda^\prime (\omega)]}{\omega+\mu-U-\lambda(\omega)}
\right ] \cr
&+& \frac{1-w_1}{\omega+\mu+i0^+}\left [ 1+\frac{(\omega+\mu)
[-1+\lambda^\prime(\omega)]}{\omega+\mu-\lambda(\omega)}\right ]\Biggr \}\cr
&+&Tw_1\ln(1+e^{\beta(\mu-U)})+T(1-w_1)\ln(1+e^{\beta\mu}).
\label{eq: contour5}
\end{eqnarray}
The terms inside \textrm{Im}\{...\} can be expressed as a derivative
\begin{eqnarray}
\mathcal{F}&=&-T\int d\omega A(\omega)\ln(1+e^{-\beta\omega})
-T\ln\mathcal{Z}+\mu_fw_1+\mu\rho_c\cr
&+&\frac{T}{\pi}\int d\omega
\ln [1+e^{-\beta\omega}]\cr
&\times&\frac{d}{d\omega}\textrm{Im}\Biggr \{ w_1\ln \left [ 1-
\frac{\lambda(\omega)}{\omega+\mu-U}\right ]\cr
&+& (1-w_1)\ln \left [1-\frac{\lambda(\omega)}{\omega+\mu}\right ] \Biggr \}\cr
&+&Tw_1\ln(1+e^{\beta(\mu-U)})+T(1-w_1)\ln(1+e^{\beta\mu}).
\label{eq: contour6}
\end{eqnarray}
Now we integrate by parts and recall Eq.~(\ref{eq: f_deriv}).  Since the
boundary terms vanish, we are left with an integral over the real axis, which
can be re-expressed in terms of the contour $C^\prime$, and then deformed
into an integral over the contour $C$.  This gives
\begin{eqnarray}
\mathcal{F}&=&-T\int d\omega A(\omega)\ln(1+e^{-\beta\omega})
-T\ln\mathcal{Z}+\mu_fw_1+\mu\rho_c\cr
&-&\frac{1}{2i\pi}\int_C d\omega f(\omega)
\Biggr \{ w_1\ln \left [ 1-
\frac{\lambda(\omega)}{\omega+\mu-U}\right ]\cr
&+& (1-w_1)\ln \left [1-\frac{\lambda(\omega)}{\omega+\mu}\right ] \Biggr \}\cr
&+&Tw_1\ln(1+e^{\beta(\mu-U)})+T(1-w_1)\ln(1+e^{\beta\mu}).
\label{eq: contour7}
\end{eqnarray}
The contour integral can be evaluated by residues which produces a sum over
Matsubara frequencies
\begin{eqnarray}
\mathcal{F}&=&-T\int d\omega A(\omega)\ln(1+e^{-\beta\omega})
-T\ln\mathcal{Z}+\mu_fw_1+\mu\rho_c\cr
&+&T\sum_n\Biggr \{ w_1\ln \left (1-\frac{\lambda_n}{i\omega_n+\mu-U}\right )
\cr
&+&(1-w_1)\ln \left (1-\frac{\lambda_n}{i\omega_n+\mu}\right )\Biggr \}\cr
&+&Tw_1\ln(1+e^{\beta(\mu-U)})+T(1-w_1)\ln(1+e^{\beta\mu}).
\label{eq: contour8}
\end{eqnarray}
The sum over Matsubara frequencies can replaced by terms that involve
$\ln\mathcal{Z}_0$ and $\ln\mathcal{Z}_1$ from Eqs.~(\ref{eq: z0_def})
and (\ref{eq: z1_def}).  Collecting terms gives
\begin{eqnarray}
\mathcal{F}&=&-T\int d\omega A(\omega)\ln(1+e^{-\beta\omega})\cr
&+&Tw_1\ln\frac{\mathcal{Z}_1}{\mathcal{Z}}+T(1-w_1)\ln\frac{\mathcal{Z}_0}
{\mathcal{Z}}+E_f w_1+\mu\rho_c
\label{eq: contour9}
\end{eqnarray}
Using the definitions for $w_0$ and $w_1$ in terms of the $\mathcal{Z}$'s,
and the relation
\begin{eqnarray}
\ln (1+e^{-\beta\omega})&=&-\beta\omega f(\omega)-f(\omega)\ln f(\omega)\cr
&-&[1-f(\omega)]\ln[1-f(\omega)]
\label{eq: f_indent}
\end{eqnarray}
gives us our final result for the Helmholz free energy
\begin{eqnarray}
\mathcal{F}&=&\int d\omega A(\omega)f(\omega)(\omega+\mu)+E_fw_1\cr
&+&T\int d\omega A(\omega)\{
f(\omega)\ln f(\omega)+[1-f(\omega)]\ln[1-f (\omega)]\} \cr
&+&T[w_1\ln w_1+(1-w_1)\ln (1-w_1)].
\label{eq: f_fk}
\end{eqnarray}
This is the Falicov-Kimball-Plischke form for the free energy which
completes the derivation. This form of the Helmholz free energy is also
correct for the Falicov-Kimball model with correlated hopping and it can
be proved in the same way starting from the expressions of
Ref.~\onlinecite{shvaika_2002b}.

The research described in this publication was made possible in part by
Award No. UP2-2436-LV-02 of the U.S. Civilian Research \& Development
Foundation for the Independent States of the Former Soviet Union (CRDF).
J.K.F. also acknowledges support from the National Science Foundation 
under grants numbered DMR-9973225 and DMR-0210717.

\bibliography{fk_dmft}

\end{document}